# A route to engineered high aspect-ratio silicon nanostructures through regenerative secondary mask lithography

*Martyna Michalska†, Sophia K. Laney†, Tao Li†, Manish K. Tiwari, Ivan P. Parkin and Ioannis Papakonstantinou[*]*

M. Michalska, S.K. Laney, T. Li, I. Papakonstantinou: Photonic Innovations Lab, Department of Electronic & Electrical Engineering, University College London, Torrington Place, London WC1E 7JE, UK

M.K. Tiwari: Nanoengineered Systems Laboratory, Department of Mechanical Engineering, University College London, Torrington Place, London WC1E 7JE, UK, and

Wellcome/EPSRC Centre for Interventional and Surgical Sciences (WEISS), University College London, London, W1W 7TS, UK

I.P. Parkin: Department of Chemistry, University College London, Torrington Place, London WC1E 7JE, UK

[†] These authors contributed equally to this work

[*] i.papakonstantinou@ucl.ac.uk

**ABSTRACT:** Silicon nanostructuring imparts unique material properties including antireflectivity, antifogging, anti-icing, self-cleaning, and/or antimicrobial activity. To tune these properties however, a good control over features size and shape is essential. Here, a versatile fabrication process is presented to achieve tailored silicon nanostructures (thin/thick pillars, sharp/truncated/re-entrant cones), of pitch down to ~50 nm, and high-aspect ratio (>10). The approach relies on pre-assembled block copolymer (BCP) micelles and their direct transfer into a glass hard mask of an arbitrary thickness, now enabled by our recently reported regenerative secondary mask lithography. During this pattern transfer, not only the mask diameter can be decreased but also uniquely increased; constituting the first method to achieve such tunability without necessitating a different molecular weight BCP. Consequently, the hard mask modulation (height, diameter) advances the flexibility in attainable inter-pillar spacing, aspect ratios, and re-entrant profiles (=glass on silicon). Combined with adjusted silicon etch conditions, the morphology of nanopatterns can be highly customized. The process control and scalability enable uniform patterning of a 6-inch wafer which is verified through cross-wafer excellent antireflectivity (<5%) and water-repellency (advancing contact angle 158°; hysteresis 1°). It is envisioned the implementation of this approach to silicon nanostructuring to be far-reaching, facilitating fundamental studies and targeting applications spanning solar panels, antifogging/antibacterial surfaces, sensing, amongst many others.

## 1. INTRODUCTION

Nanopatterning of silicon to afford superior functionality and performance has become commonplace, with demonstrations spanning solar cells,[1,2] energy storage,[3] thermoelectrics,[4] sensors,[5] antibacterial[6,7] and special wetting surfaces.[8–10] To elicit the desired and new functionalities however, it is critical to control both the morphology and dimensions of nanofeatures while accommodating a continuous demand for higher resolution and aspect ratio (AR). Such requirements concurrent with method scalability, reliability, and compatibility with existing manufacturing processes are not trivial to achieve, bringing a necessity to advance nanofabrication techniques. In this regard, pattern transfer using block copolymers (BCPs) has been investigated due to their low cost, morphological diversity and proficiency in high resolution patterning (5-200 nm);[11–13] overcoming the limitations posed by photolithography and electron beam lithography, for instance.

Nonetheless, BCP lithography holds its own challenges related either to feasibility of morphology control or insufficient etching contrast. To overcome the former, substrate preparation (neutral brush layer), thermal/solvent annealing, and development steps have been adapted.[14–16] Here, conditions need to be carefully chosen, particularly during annealing to prevent de-wetting or undesired morphology formation.[17] This increases the complexity and cost of the fabrication, which constitutes a potential challenge for implementation in industry. Furthermore, control over pitch demands individual optimization for BCPs of different molecular weight $M_w$, limiting flexibility. Recently, an alternative BCP micelle lithography process was presented whereby pre-assembled solution-phase micelles of poly(styrene – *block* – 2-vinylpyridine) (PS-*b*-P2VP) were directly spin-coated onto a substrate, yielding micellar bumps which act as a topographic contrast.[18,19] This route negates the aforementioned steps, whilst enabling reduction of mask diameter, pitch fine-control through spin speed variation, and coarse-control through choosing a different $M_w$ BCP (without individual optimization).

To achieve high AR nanostructures, a large etching contrast is required. Here, one common route is choosing a silicon-containing BCP – such as PDMS – which under oxygen plasma converts into silicon oxycarbide, leaving a hard mask with enhanced mechanical and thermal stability.[20] This allowed for fabrication of silicon nanopillars/nanoholes;[17,21] yet with an AR<2 due to the limited BCP thickness, therefore necessitating an additional layer such as chromium.[22] Alternatively, the constituent blocks of organic BCPs can be selectively infiltrated with metals/metal oxides,[23–25] to yield high AR arrays of nanopillars/cones/gratings. In these techniques however, the experimental conditions are delicate, demanding careful matching of precursor and block chemistry to obtain good infiltration efficiency, and in some instances requiring expensive equipment (atomic layer deposition). The cost and impracticality therefore represent a key limitation, with the additional potential introduction of metal contaminants into the chamber. Instead, using a thin 20-25 nm intermediate $SiO_2$ layer as a hard mask has resulted in high AR porous nanostructures (AR~10),[26] albeit with limited success and control for more challenging geometries like pillars and cones; with the achieved AR<2, and some control over morphology.[18]

Herein, we present a library of precisely tailored Si nanostructures originating from BCPs (pillars, cones, and re-entrant) with pitches ranging from ~50-260 nm. Using pre-assembled BCP micelles, we apply our recently reported regenerative secondary mask lithography (RSML) process[27] to enable the pattern transfer into an intermediate $SiO_2$ layer (hard mask) of an arbitrary thickness. Not only does this method solve the etching contrast problem but it also uniquely allows the hard mask diameter to be increased as well as decreased without necessitating a different $M_w$ BCP. Through modulation of the hard mask, alongside the Si etch conditions ($Cl_2$ flow, coil/platen power), high AR (>10) Si nanostructures can be generated with precisely tuned morphologies, permitting rational design. Finally, as a proof-of-concept, we demonstrate on a 6"-wafer scale that by engineering the surface nanostructures, an excellent antireflective and robust superhydrophobic surface with ultralow hysteresis (1º) is obtained.

## 2. RESULTS AND DISCUSION

**Fabrication**. A schematic of the strategy to fabricate Si nanostructures is shown in Figure 1 and it relies on BCP transfer into the $SiO_2$ hard mask, followed by pattern registration in the underlying Si. An $SiO_2$ layer of thickness $T$ corresponding to the desired height $h$ of the hard mask is first deposited on a silicon wafer. Subsequently, micelles of PS-*b*-P2VP are pre-assembled through dissolution in *m*-xylene and spin-coated to generate hexagonally-packed micellar bumps (step 1; Figure 1), as previously described.[18] The center-to-center distance (pitch, *p*) is determined predominantly through the $M_w$ of each block, the solvent, and spin speed; and it ranges here from 56-257 nm as shown in scanning electron microscopy (SEM) images in Figure 1, and Figure S1. A brief and mild oxygen breakthrough etch is performed to expose the underlying $SiO_2$ by removing the PS matrix (step 2). This step additionally provides an opportunity to decrease the mask diameter *d* but at the cost of its height *h*, lowering the etching contrast in classical approaches. However, the implementation of the RSML for $SiO_2$ etching in the third step overcomes this issue through inducing secondary mask formation, which acts as a protective layer enhancing BCP durability; hence enabling the pattern to be uniformly transferred through the entire thickness of the deposited $SiO_2$ layer. This allows for not only $SiO_2$ height control but also diameter control by varying $H_2$ content in the gas feed ($CHF_3/Ar/H_2$) so that, simply put, the higher the $H_2$ amount, the greater the mask diameter, as shown schematically in Figure 1 (step 3). Further details are provided in the proceeding sections, where we demonstrate $SiO_2$ mask of $h$ = 350 nm corresponding to an AR>3.

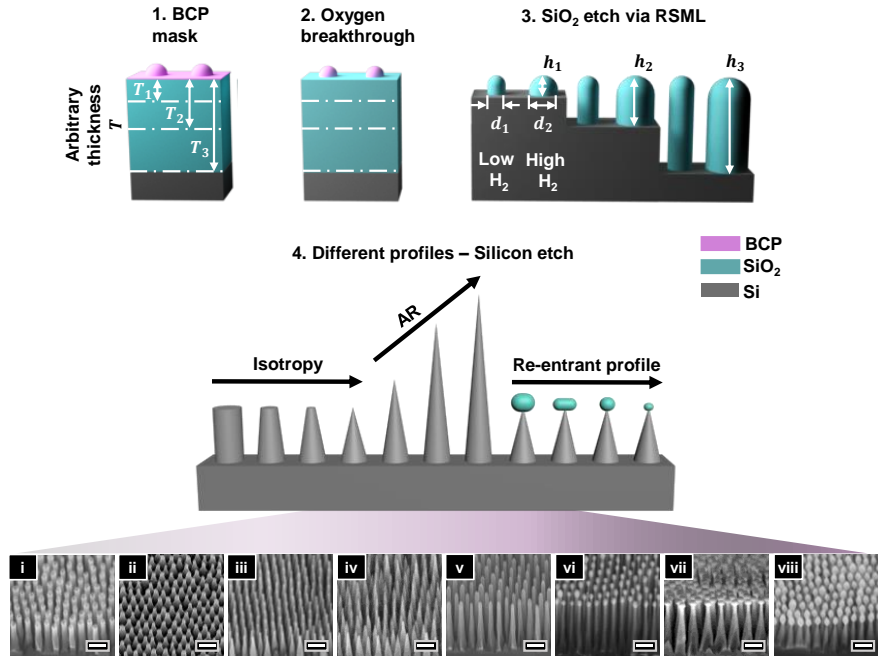

**Figure 1.** Schematics of the four key process steps to generate a range of tailored Si nanostructures. (a) Mask preparation (1-3) and pattern transfer (4). (1) Spin-coating of pre-assembled micelles to generate hexagonally-packed arrays with typical height $h\sim20$ nm; the substrate consists of Si with an $SiO_2$ layer of pre-determined thickness $T$; (2) Isotropic oxygen breakthrough etch to remove the PS matrix. Note, the diameter/height ($d/h$) of the bump can also be reduced at this stage; (3) Transfer of pattern into $SiO_2$ hard mask, where addition of $H_2$ gas into the etching chemistry protects the BCP mask through increased carbopolymer deposition. This enables the generation of hard masks with high ARs ($h$ dependent on the $T$ of deposited $SiO_2$ layer) and permits the $d$ of the mask to be increased ($d_1<d_2$). (4) Etching into Si leads to a library of morphologies: with varying sidewall angle [isotropy; straight-walled pillars (**i**) and cones (**iv**)]; degree of truncation upon the mask removal (**ii-iv**); aspect ratio AR (**iv-vi**), as well as re-entrant profile (**vi-viii**). The pitch is 56 nm (**i**) and 110 nm (**ii-viii**). The end Si structures are engineered based on the $h$ and $d$ of $SiO_2$ mask, in addition to the etching conditions (coil and platen power/pressure/time). Scale bars are 200 nm.

The pattern transfer into the Si layer (Figure 1; step 4) is governed by a combination of the $SiO_2$ mask morphology ($h/d$), and the Si etch conditions. Here, we use $Cl_2$ plasma, with a characteristic etching selectivity >5, dependent on the coil/platen power, $Cl_2$ flow, and pressure. Through manipulation of the glass and Si etching, we realize high AR nanostructures with sophisticated sidewall profiles, including nanopillars, nanocones with truncated or sharp tips, re-entrant structures and nanopyramids, all of which are discussed separately in the following sub-sections.

**Tuning the $SiO_2$ mask through RSML**. Typically glass nanostructuring via RIE proceeds through a combination of; (i) chemical etching with a fluorocarbon plasma such as $CHF_3$, (ii) ion-assisted etching with an inert species such as argon, and (iii) simultaneous fluorocarbon deposition $CF_x$. However, as the etching chemistry stands ($CHF_3/Ar$), the deposition is not sufficient to prevent premature consumption of the non-robust organic BCP mask, thus culminating in very low

AR of SiO$_2$ masks. Nonetheless, the addition of H$_2$ increases the formation of HF, in turn lowering the F/C ratio and generating a more polymerizing plasma (greater CF$_x$ deposition). At such conditions, the polymer build-up at the top of the structure can be induced,[27] embedding the BCP within a secondary organic mask; depicted schematically in Figure 2a,ii and shown in the SEM inset of Figure 2b. However, left unattended, over-deposition can block the path of bombarding ions and etching species, and prevent further etching. Therefore, in order to attain structures of higher AR, we apply a brief oxygen plasma which acts to controllably reduce the size of the secondary mask, allowing further etching at the base to proceed (Figure 2a,iii). During the subsequent etch, the mask regenerates as described in our previous work on patterning fused silica.[27] This cycle of etching followed by an oxygen breakthrough can be repeated numerous times to reach the desired AR. In Figure 2b, we demonstrate the achieved SiO$_2$ structures after 2 cycles with $p = 110$ nm, and $h = 350$ nm.

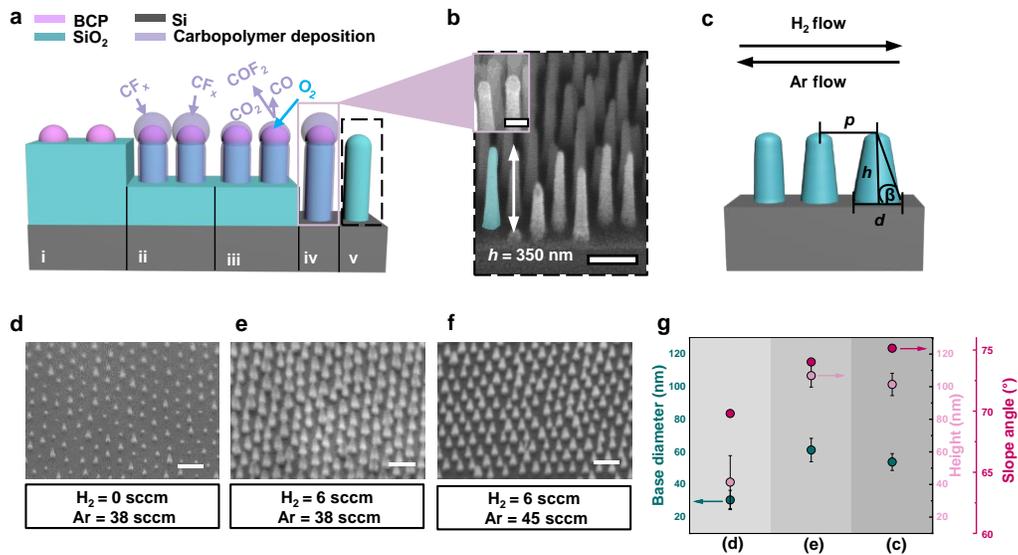

**Figure 2.** Tuning of the SiO$_2$ hard mask height and shape. (a) Schematic representation of the SiO$_2$ nanopillar generation through RSML. Starting from the BCP (i), the addition of H$_2$ into the etching chemistry (CHF$_3$/Ar) increases the generation of HF, thus lowering the F/C ratio; resulting in a more polymerizing plasma with increased CF$_x$ deposition (ii). As the secondary mask builds up around the BCP to form mushroom-like structures, a brief oxygen plasma is applied which refines the diameter (iii). This enables further etching to proceed, with regeneration of the secondary mask to yield tall SiO$_2$ nanopillars (iv). A final oxygen plasma removes the deposition to form high AR SiO$_2$ nanopillar masks (v). (b) SEM image of the generated hard mask with $h = 350$ nm, and $p =1$ 10 nm (AR>3). The inset shows the structure as depicted in (iv) with polymer deposition remaining. Scale bar (inset) = 100 nm. (c) Schematic representation of the effect of increasing H$_2$ or Ar flow, with key parameters depicted; height $h$, base diameter $d$, slope angle $β$, pitch $p$. (d-f) SEM images of SiO$_2$ hard masks generated under varying H$_2$ and Ar flows with an initial SiO$_2$ thickness of 100 nm. No H$_2$ flow results in partial mask destruction with uneven topography (d). Maintaining the same Ar flow but increasing H$_2$ flow yields taller, more uniform SiO$_2$ pillars with a wide base diameter (e). Maintaining the same H$_2$ flow but increasing Ar flow, gives rise to similarly tall and uniform SiO$_2$ pillars, but with a narrow base diameter (f). (g) Corresponding

quantitative analysis of the change in base diameter, height and slope angle for the structures shown in SEM images d-f. Scale bars = 200 nm.

Not only does the altered etching chemistry permit high ARs through increased selectivity, but additionally enables the mask profile to be tuned so that the desired base diameter and anisotropy can be attained in order to well control the structure generation in Si. Profile control relies on the precise adjustment of $H_2$ and Ar flow, shown schematically in Figure 2c and in a series of SEM images in Figures 2d-f. Without $H_2$ ($CHF_3$:Ar = 1:3.2), the mask erosion quickly occurs resulting in non-uniform pattern transfer with a large *h* distribution, a small *d*, and low slope angle *β* (Figure 2d); as measured and quantitatively represented in Figure 2g. Introducing $H_2$ so that $CHF_3$:$H_2$:Ar = 2:1:6.3 (Figure 2e), yields 2.5-fold taller structures with a narrow distribution, twofold greater *d*, and larger *β*. Alternatively, achieving more anisotropic structures (even larger *β*) with a reduced *d* whilst retaining maximum *h*, is possible by decreasing carbopolymer deposition via stronger physical bombardment (increased Ar flow), so that $CHF_3$:$H_2$:Ar = 2:1:7.5 (Figure 2f).

**Tuning the Si nanostructures.** We next utilize the hard masks of varying diameters and heights, obtained from the same BCP template through modulation of RSML conditions, to elicit a range of designer nanostructures in silicon (Figure 3). For example, in Figure 3a, two generated hard masks of *d* = 60 and *d* = 112 nm, yield nanocones with ultra-sharp tips (i) and straight-walled nanopillars (ii) when the same etching conditions are applied ($Cl_2$ plasma; moderate power). It is noteworthy that ultra-sharp tips are often formed by thermal oxidation, followed by removal of the oxide layer.[28] Bypassing these steps therefore decreases process complexity. Whilst precisely controlled nanocone/nanopillar arrays present surfaces with invaluable properties (e.g., antireflective,[29] antifogging,[10] antibacterial[30]), unconventional morphologies, such as re-entrant structures, have been shown to display extraordinary omniphobic properties through manipulation of the direction of the liquid-vapor interface.[31] Starting from the same $SiO_2$ mask, but applying a reduced coil (<200 W) and particularly platen power (<15 W), we achieve mask undercutting which gives rise to nanopyramids of different re-entrant profiles (Figure 3a(iii, iv)). Here, the mask is observed to remain at the top of the structure and can be either stripped away or depleted through further etching leading to slender pillars of higher AR (Figure S2). More details on controlling the re-entrant profile are discussed in the following sections.

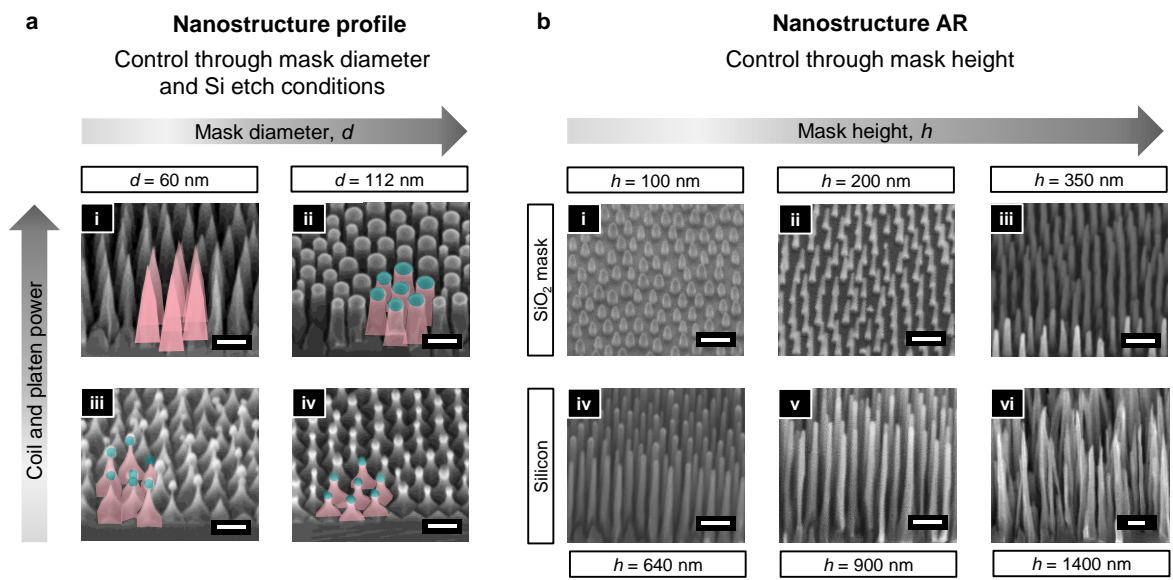

**Figure 3.** Tuning of nanostructure profile and aspect ratio. SEM images presenting an effect of the hard mask diameter (a) and height (b) on the pattern transfer into Si. (a) Control over the nanostructure profile is obtained through a combination of the $SiO_2$ hard mask dimensions and the applied Si etching conditions (coil/platen power). Originating from the same BCP, Si nanocones (i) and straight-walled pillars (ii) were generated upon etching of a narrow ($d = 60$ nm) and wide ($d = 112$ nm) $SiO_2$ mask, respectively; and under a coil power of 300 W and platen power of 40 W. Nanopyramids (iii and iv) were similarly generated upon etching of the narrow and thin mask, but under a reduced coil (200 W) and platen (10 W) power. Both nanopyramids are similar, but the mask remaining in (iii) is very fragile and can be easily removed through HF treatment to yield short and sharp nanocones, whereas (iv) possesses considerable remaining hard mask. (b) Control over the Si nanostructure's AR is provided through RSML-generated $SiO_2$ masks of varying height. $SiO_2$ masks of height 100 nm (i) and 200 nm (ii) were generated within one RSML cycle, whereas the mask of height 350 nm (iii) required two cycles. Etching of masks (i-iii) under coil power 300 W and platen power 40 W yields nanostructures (iv-vi) of heights 640, 900, and 1400 nm, respectively. Scale bars = 200 nm.

As in glass etching, a facile route for high aspect ratio silicon nanostructure generation has been absent, due to the lack of BCP mask durability. However, through RSML, we increase the AR of the $SiO_2$ hard mask, and overcome this issue. Figure 3b shows three hard masks of increasing $h$ (100-350 nm), and the corresponding Si nanostructures etched under the same conditions. All structures are etched to the point of mask consumption to yield the highest AR (AR>10, Figure 3,vi), however this can introduce distortion. To avoid this therefore, etching should be stopped just prior to hard mask depletion, and followed by post-processing hard mask removal (e.g., HF treatment). Important to note, is that high AR hard masks, such as that shown in Figure 3b,iii, can lead to bowing as a result of the mask weight and the AR of the Si structures. Nonetheless, the resulting Si nanograss (Figure 3b,vi) are likely to bring antireflective properties[32] or may be effective in antibacterial performance.[33]

The nanostructure design can also accommodate geometries with different levels of truncation, as highlighted in the SEM images in Figure 4. Figure 4a,i is the most truncated, with nanostructures displaying great similarities to those found on the wings of the Cicada.[34] Conversely, Figure 4a,iv possesses no truncation. Here, Si etching occurs in the same manner as the re-entrant structure generation (reduced coil (200 W) and platen (10 W) power), however it is stopped before complete undercutting of the hard mask occurs. The degree of truncation, therefore, is dictated by the etching time, with the top diameter of the truncated cone corresponding directly to the base diameter of the remaining hard mask. The inset SEM images of Figure 4a show the structures before HF treatment; which is required to remove the remaining hard mask to attain truncated or sharp tops.

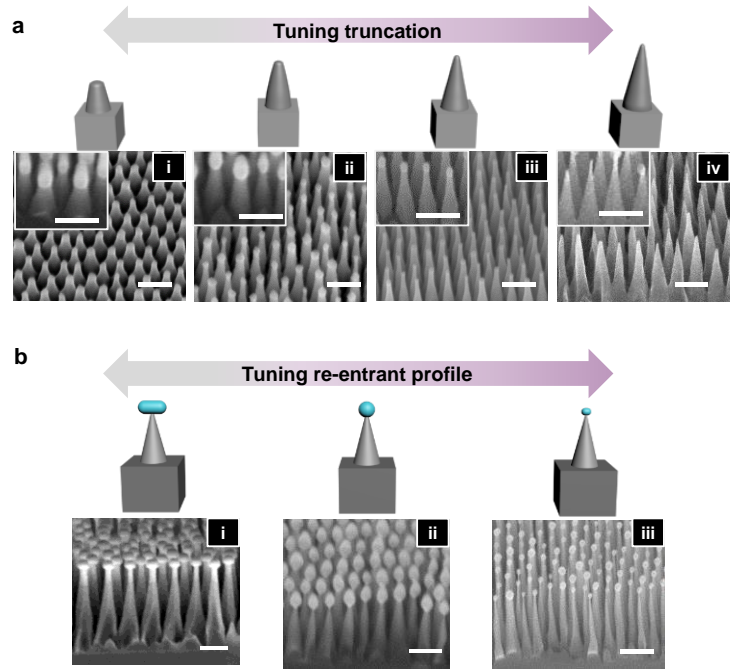

**Figure 4.** Tuning of truncation and re-entrant profile. Si nanostructure morphology. (a) SEM images and schematics of Si nanostructures with varying degrees of truncation –  high (i) to low (iv) – generated by premature etch stop with subsequent HF treatment to remove the hard mask (inset shows structure before mask removal). The application of different etching time under the same isotropic etching conditions results in variable degrees of nanocone truncation. (b) Schematics and SEM images of the re-entrant nanostructures obtained under the same etching conditions but originating from a wide and flat (i) or a taller and tapered (ii) hard mask. (iii) Schematic and SEM image of a re-entrant nanostructure with a greater height and reduced hard mask generated from the same mask as in (ii) but under different etching conditions (more directional plasma). The application of the same (different) isotropic etching conditions to different (the same) shapes of the hard mask results in variable re-entrant profile. Scale bars = 200 nm

Controlling the re-entrant profile is also possible as shown in Figure 4b, and it relies predominantly on the dimensions of the hard mask. For example, Figure 4b,i originated from a flat and wide hard mask, whereas Figure 4b,ii originated from a taller and rounder hard mask. Nonetheless, the etching conditions and time also play a role. Here, Figure 4b,iii originating from the same mask as Figure 4b,ii was etched under higher coil/platen power and for a shorter time, yielding taller morphologies with less remaining hard mask. The slope angle $β$ of the Si nanostructures can be tuned in a similar manner through control of the hard mask dimensions (shorter hard mask can elicit shorter, nanocones with a smaller $β$ and *vice versa*), and through the Si etching conditions (Figure S3).

**Applications.** The wealth of potential morphologies which can be reliably attained here unlocks many functionalities through the ability of the nanostructure to manage the interactions with liquids, photons, and bacteria.[2,6,35] As an example, we therefore investigate here the antireflectivity and superhydrophobicity of nanostructured silicon whilst presenting the method scalability by patterning the surface of a 6"-wafer. Figure 5a shows a large-area SEM image of the nanostructures clearly showcasing the pattern uniformity and a photograph demonstrating a uniform black color across the entire surface, with no visible reflections. To further demonstrate that there is consistent antireflectivity (also an indication of structural homogeneity) across the wafer, we measure the reflectance at the 5 locations marked in Figure 5b and plot the absolute reflectance alongside that of the unstructured Si over the wavelength range 400-1000 nm (Figure 5c). This further demonstrates the quality of the pattern with an average reflectance of <5% and only slight variation across the wafer (<1%); meanwhile, the unstructured surface exhibits an average reflectance >40%. To render the wafer superhydrophobic, a thin layer of short-chain PDMS is grafted onto the surface,[36] and the wetting properties characterized by dynamic water contact angle measurements. Excellent superhydrophobicity is observed with high advancing (158 ± 1°) and receding (157 ± 1°) contact angles and an ultralow contact angle hysteresis (1°). Such high water-repellence enables droplets to bounce off a surface multiple times with minimal dissipation of energy, and we observe a remarkable 19 bounces upon the release of water droplet (8 µl) from 1 cm height (Figure 5d). Furthermore, we demonstrate an excellent self-cleaning performance, where black pepper powder is efficiently removed by a stream of water.

## 3. CONCLUSIONS

In summary, we have developed a fully tunable and simple fabrication route for a wide range of silicon nanostructure morphologies that vary in aspect ratio and shape at the length scales challenging to reach by conventional techniques. These multiple nanostructures originate from only one type of BCP micelle template, which we modulate further by RSML process so that various $SiO_2$ hard mask heights (>300 nm) and diameters (increase or decrease with respect to the original micelle) become available. The flexibility of the method stemming from numerous permutations of $SiO_2$ and Si etching conditions offers an easy to adapt platform for both Si nanopatterns as well as generation of $SiO_2$/Si heterostructures. Finally, we demonstrate the process scalability through patterning of 6"-wafer with nanopillars that can be utilized as highly antireflective and self-cleaning substrate. We envision the implementation of our approach to Si nanostructuring to be far-reaching, targeting applications which require large-scale, uniform patterning such as solar panels, anti-fogging/anti-bacterial surfaces, sensing, amongst many others.

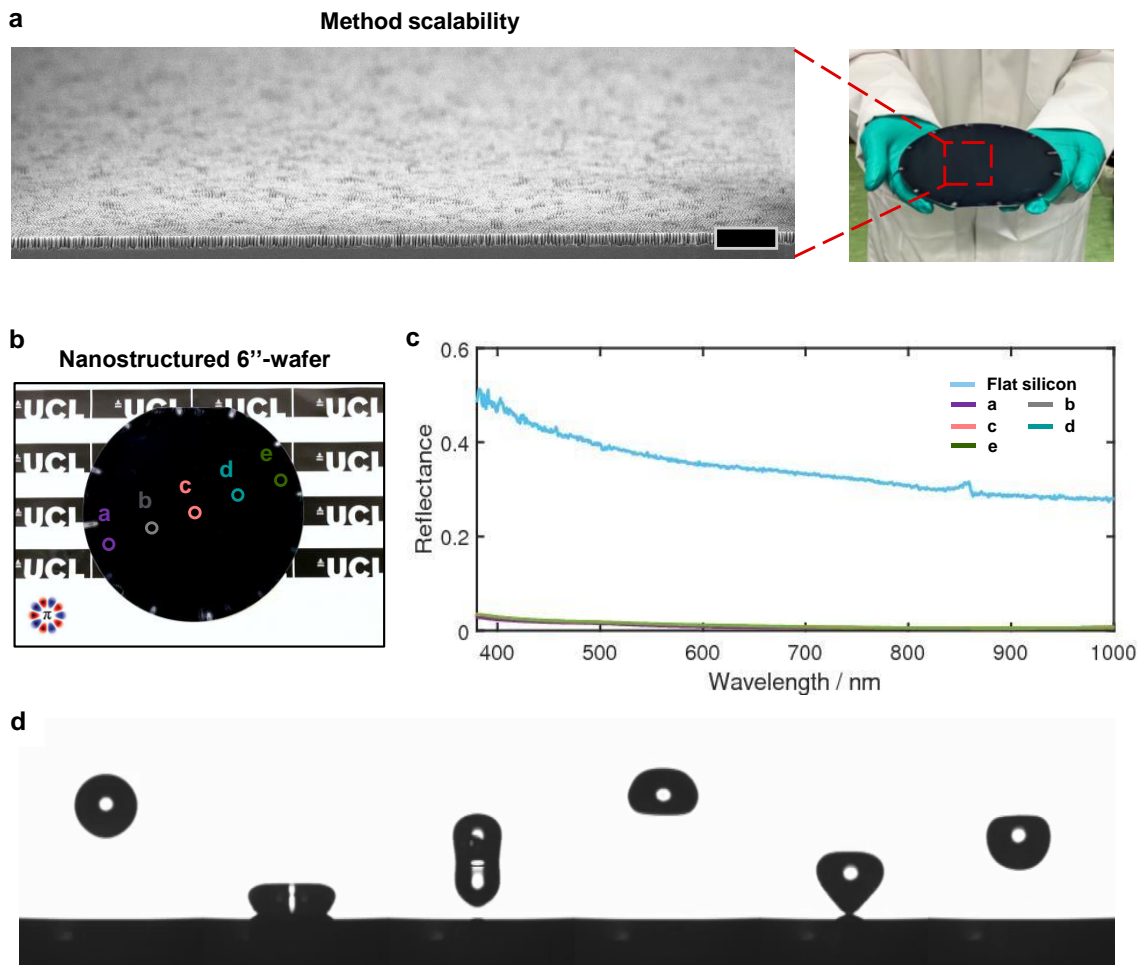

**Figure 5.** Method scalability, optical properties, and wetting characteristics of the nanostructured silicon wafer. (a) Large field of view SEM illustrating silicon nanostructures of AR~7 of the 6''-wafer (photograph), which demonstrates the scalability of the approach. Scale bar is 2 μm. (b) Photograph of the nanostructured 6''-silicon wafer with the five locations marked where reflectance measurements were taken. (c) Measured reflectance as a function of wavelength for flat silicon (control) and the nanostructured silicon at the five locations marked in (b). (d) Sequential images of a droplet impacting the structured surface, with minimal dissipation of energy.

## 4. EXPERIMENTAL SECTION

*Fabrication of silicon nanostructures.*

*SiO$_2$ deposition.* A silicon wafer (MicroChemicals) was cleaned with acetone via sonication and subsequently washed with isopropanol. An SiO$_2$ layer of thickness 40, 70, 100, 200, or 350 nm was deposited on Si via plasma enhanced chemical vapor deposition (PECVD; STPS Multiplex) under low frequency RF with SiH$_4$ and O$_2$ vapor at 300°C.

*BCP micelle preparation.* BCP micelles of PS-b-P2VP [poly(styrene-block-2vinylpyridine), Polymer Source Inc.] were pre-assembled according to the previous report,[18] with certain adaptations. Three molecular weights were used for this study Mw / kg mol-1 = 57-b-57 (P57), 109-b-90 (P100), and 440-b-353 (P400) to accommodate pitches ranging from ~50-250 nm. The polymers were mixed with anhydrous m-xylene at concentrations of 0.3-0.5% w/v by gentle stirring at 75°C overnight. Subsequently, the solutions were allowed to cool to room temperature (RT), filtered (PTFE 1 µm), and stored at 4°C.

*SiO$_2$ hard mask preparation.* First, pre-assembled micelles were spin-coated onto the Si/SiO$_2$ wafer at RT. Typical spin speeds for P400 ranged between 2-4k rpm, and for P100 and P57: 3-6k rpm. To register the pattern into SiO$_2$, reactive ion etching (RIE) was conducted using PlasmaPro NGP80 RIE, Oxford instruments, at temperature of 20°C. A breakthrough etch (3-14 s) was performed to remove the PS matrix and tune the diameter of the mask under O$_2$ (20 sccm), pressure 50 mTorr, and radio frequency (RF) power 50 W. Subsequently, glass was etched according to our recently reported RSML process[27] using CHF$_3$/H$_2$/Ar gases at flows 12-15, 0-6, and 38-45 sccm, respectively; under pressure of 30 mTorr and at RF power of 220 W. Control over the etching depth was obtained through the time, and to reach high aspect ratio, a breakthrough etch was performed under O$_2$ plasma (conditions as above) to reduce the diameter of the secondary mask. Etching of SiO$_2$ proceeded until reaching the underlying Si layer, at which point an O$_2$ plasma clean was performed to remove the organic mask.

*Si etching.* The pattern from the hard mask was transferred into Si by means of an Advanced Silicon Etcher (ASE, STS MESC Multiplex ICP) under Cl$_2$ plasma. The conditions varied depending on the desired degree of anisotropy, with an anisotropic etch performed under coil power 300 W and platen power 40 W with Cl$_2$ flow of 20 sccm and pressure 3 mTorr. To obtain a less anisotropic profile (re-entrant), the etch was performed under a coil power of 150-200 W and platen power 10-15 W, with a Cl$_2$ flow of 20 sccm and pressure 3 mTorr. In some cases, a Cl$_2$ was mixed with SF$_6$ to increase the degree of hard mask undercutting (see details in Figure S3).

*Surface functionalization.* In order to render surface superhydrophobicity, a silanization was performed. The substrate was first cleaned via sonication in isopropanol and acetone, and subsequently treated with oxygen plasma to impart surface hydroxylation (5 min each). Short chain polydimethylsiloxane (PDMS) was grafted onto the surface, as previously reported,[36] using a 1:10:0.27 v/v/v ratio of dimethyldimethoxysilane : isopropanol : H$_2$SO$_4$ (>95%) mixture. The substrate was then placed on a hot plate (75°C) and the solution was drop casted atop for 15 s, followed by washing with deionised water, isopropanol, and toluene.

*Surface characterization.* Atomic Force Microscopy (AFM). The BCP micelle patterns were evaluated using an AFM (Dimension Icon-PT from Bruker AXS) in tapping mode at room temperature. The pitch was determined using ImageJ (https://imagej.nih.gov/ij/) software with the nearest neighbor distance plugin. Scanning electron microscopy (SEM). The SEM images were taken by a Carl Zeiss XB1540 SEM at 2-5 kV operating voltage with a tilt angle of 45° or 90°. ImageJ was used for statistical analysis of the nanostructure dimensions such as pitch, height, diameters.

*Wettability.* The advancing and receding contact angles were measured using a custom designed goniometry set up. The setup consists of syringe pump (Cole-Parmer Single-syringe infusion pump), a needle (BD PrecisionGlide™ needles, 21G), and an imaging device (Thorlab, model

DCC1240). Droplets of approximately 30 µl were deposited onto the surfaces and further extracted using the syringe pump to measure advancing and receding contact angle, respectively. The videos taken during droplet deposition and extraction were processed through a Matlab script for contact angle measurements,[37] which is available from the corresponding author upon reasonable request. Droplet bouncing was characterised by releasing an 8µL droplet from a height of 1cm. A high-speed camera (Phantom V411 fitted with a macro lens) was used to record and count the number of bounces.

*Reflectance.* Reflection measurements were taken at 8° off normal incidence. The sample was attached to a port of an integrating sphere (Labsphere) and illuminated using a white light source (KI-120 Koehler Illuminator, Labsphere). Light levels were measured using a fiber coupled spectrometer (QEPro, Ocean Optics) and calibrated against a diffuse reflectance standard (SRS-02-10, Spectralon, Labsphere).

**Supporting Information**
The Supporting Information accompanies this manuscript and contains Figures S1-S3.

**Acknowledgments**
This work was conducted in the framework of the European Research Council, ERC-StG-IntelGlazing, grant no: 679891. S.L. and I.P. thank UKRI/EPSRC for a DTP award grant no EP/N509577/1 and funding from Lloyd's Register Foundation International Consortium of Nanotechnology (ICON) research grant. MKT acknowledges the Royal Society Wolfson Fellowship and the NICEDROPS project supported by the European Research Council (ERC) under the European Union's Horizon 2020 research and innovation programme under grant agreement no. 714712. Finally, we acknowledge the assistance of the technical team in the London Centre for Nanotechnology (LCN).

**Conflict of Interest**
The authors declare no competing financial interests.

# Supporting Information

# A route to engineered high aspect-ratio silicon nanostructures through regenerative secondary mask lithography

*Martyna Michalska,[†] Sophia K. Laney,[†] Tao Li,[†] Manish K. Tiwari, Ivan P. Parkin and Ioannis Papakonstantinou[*]*

[†] These authors contributed equally to this work

[*] i.papakonstantinou@ucl.ac.uk

## Supporting Figures

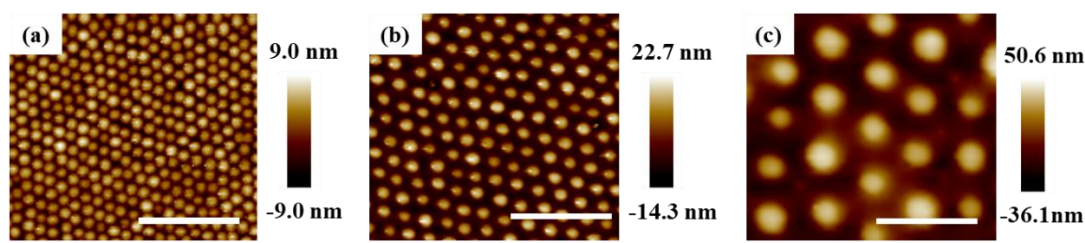

**Figure S1.** AFM images of hexagonally packed PS-*b*-P2VP micellar bumps generated from direct spin coating onto Si/SiO$_2$. The molecular weight $M_w$ dictates the pitch $p$ of the micelles and varies across (a-c) with $M_w$ / kg mol$^{-1}$ = 57-*b*-57, 109-*b*-90, and 440-*b*-353, giving rise to $p$ = 56, 95, and 257 nm, respectively. Scale bars = 500 nm.

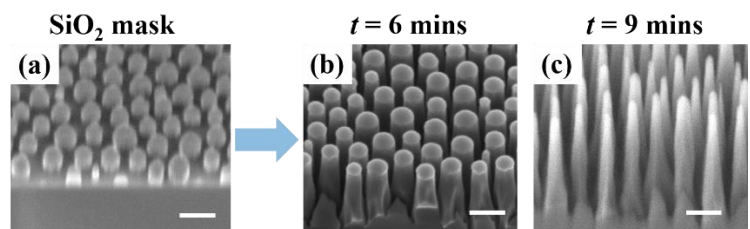

**Figure S2.** SEM images of the structures generated from a hard mask with $d$ = 112 nm. (a) SEM image of the hard mask, which under coil power 300 W and platen power 40 W yields straight walled nanopillars (b) with mask remaining after 6 min. Etching for a further 3 min under the same condition yields slender high AR nanopillars. Scale bars = 200 nm.

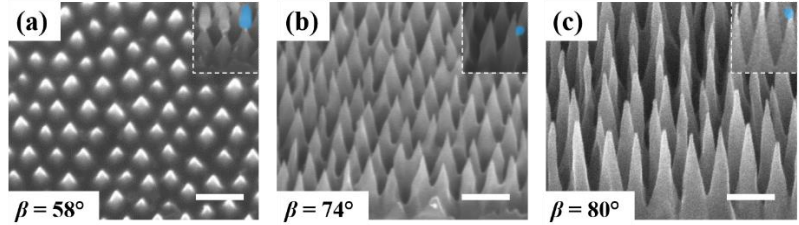

**Figure S3.** SEM images of sharp nanocones with varying slope angles after HF treatment (inset = before HF treatment, with remaining $SiO_2$ hard mask highlighted in blue). (a) Nanocones with $\beta = 58°$, generated from a hard mask of $h = 70$ nm under mixed plasma of $Cl_2$ (18 sccm) and $SF_6$ (2 sccm) at coil power 300 W, and platen power 10 W for 3 min. (b) Nanocones with $\beta = 74°$, generated from a hard mask of $h = 40$ nm under $Cl_2$ plasma (20 sccm) at coil power 200 W, and platen power 10 W for 25 min, followed by 1 min etching under mixed plasma: $Cl_2$ (19 sccm) and $SF_6$ (1 sccm). (c) Nanocones with $\beta = 80°$, generated from a hard mask of $h = 100$ nm under $Cl_2$ plasma (20 sccm) at coil power 200 W, and platen power 15 W for 15 min. Scale bars = 200 nm.